\documentclass[floats,aps,amsmath,showpacs,twocolumn]{revtex4}
\usepackage{amsmath}
\usepackage{graphicx}
\usepackage{amsfonts}
\newcommand{\tE}{t_{\mathrm{E}}}
\newcommand{\bq}{{\bf q}}
\newcommand{\bp}{{\bf p}}
\newcommand{\bx}{{\bf x}}
\newcommand{\by}{{\bf y}}
\newcommand{\Eav}{E_{\rm av}}
\newcommand{\vd}{{\vphantom{\dagger}}}
\begin{document}

\title{Universal spectral correlations from the ballistic sigma model}
\author{Jan M\"uller, Tobias Micklitz, Alexander Altland}

\address{Institut f{\"u}r Theoretische Physik, Z{\"u}lpicher Str 77, 50937 K{\"o}ln,
  Germany}

\begin{abstract}
  We consider the semiclassical ballistic $\sigma$-model as an
  effective theory describing the quantum mechanics of classically
  chaotic systems. Specifically, we elaborate on close analogies
  to the recently developed semiclassical theory of quantum
  interference in chaotic systems and show how semiclassical 'diagrams' involving
  near action degenerate sets of periodic orbits emerge in the field
  theoretical description. We further discuss how the universality phenomenon (i.e. the
  fact that individual chaotic systems behave according to the
  prescriptions of random matrix theory) can be understood from the
  perspective of the field theory.
\end{abstract}

\maketitle

\section{Introduction}
\label{sec:introduction}

In recent years, significant progress has been made in understanding the
semiclassical basis of universality in quantum
chaos. In
a sequence of steps~\cite{aleinerlarkin,sieber_richter01,smueller1,smueller2,smueller3,aa+haake}, the theory of action correlations and periodic
orbits has advanced to a stage, where the full information stored in
the spectral correlation functions of random matrix theory (RMT) has
become accessible by semiclassical methods. An important element in
this development has been the establishment of a cross-link between
semiclassics and the so-called zero dimensional sigma model (which in
turn had long been known~\cite{efetov} to be in one-to-one
correspondence to RMT): under conditions of global hyperbolicity and
for low energies (energies lower than the inverse $\hbar/t_E$ of the
so-called Ehrenfest time, $t_E$), individual contributions to the
semiclassical expansion of spectral correlation functions become fully
universal in that they depend on combinatorics and topology of the
underlying orbit correlations, but not on system specific
details. Each such term has a corresponding contribution to the loop
expansion of the zero dimensional sigma model around one of its saddle
points. From a certain perspective, the semiclassical analysis of
spectral correlations has, thus, been tantamount to a reduction of the
full microscopic information stored in the Green functions of an
individual chaotic system down to the core data encapsulated in the
zero dimensional sigma model.

In this paper, we discuss an alternative reduction scheme, which is
similar in spirit, but methodologically different. Our starting point
will be the observation that the zero dimensional $\sigma$-model
affords an alternative interpretation, viz. as the globally uniform
mean field limit of the ballistic $\sigma$-model.  The ballistic
$\sigma$-model in its variant considered here is an effective field
theory of chaotic systems, obtained from a microscopic parent
theory~\cite{bsm,ASAA1,ASAA2} by projection onto the sector of fields
fluctuating on length scales $\gtrsim \hbar$. We will present various
arguments to the effect that this projection captures the essential
physics of the problem. The resulting effective field theory is
defined on shells of conserved energy in classical phase space. Its
dynamics is controlled by the -- equally classical -- Liouville
operator. Quantum mechanics enters the theory through a feature known
as 'non-commutativity'. In practice, non-commutativity implies that
points in phase space can be defined only up to a maximal resolution
set by the Planck cell. (Notice that the typical linear extensions of
Planck cells are of $\mathcal{O}(\hbar^{1/2})$, much bigger than the
cutoff length of the theory.) Below we will explore the conspiracy of
classical hyperbolicity and Planck cell resolution in the long time
dynamics of chaotic systems.

Specifically, we will show that a perturbative expansion of the model
leads to structures that are remarkably similar, if not equivalent, to
those arising in periodic orbit theory: taking the first quantum
correction to the spectral correlation function (the 'Sieber-Richter
term'~\cite{sieber_richter01}) as an example, we will recover
information on the phase space available to individual families of
periodic orbits, and the corresponding action correlations. The actual
integrals describing these corrections turn out to be identical to
those of periodic orbit theory. This finding suggests a quantitative
identification of the Feynman diagrams of the theory with the
action-correlated orbit pairs of semiclassical analysis (However, an
extension of the calculation to higher orders in perturbation theory,
necessary to substantiate this claim, is beyond the scope of the
present paper.)

Going beyond perturbation theory, we will demonstrate that for
correlation energies smaller than the inverse of the Ehrenfest time
non-uniform fluctuations in phase space are effectively
suppressed. The resulting theory of uniform fluctuations (viz. the
zero dimensional $\sigma$-model) predicts universal behavior, in
agreement with the predictions of RMT. Field theory 
thus provides a comparatively simple way of understanding the basis
of the Bohigas-Gianonni-Schmid~\cite{BGS} universality hypothesis,
alternative to the explicit
summation over infinite hierarchies of periodic orbits~\cite{aa+haake}.
The computational efficiency of the 'mean
field plus fluctuations' approach to universality arguably represents
the most important advantage of the present formalism.

The rest of the paper is organized as follows: in section
\ref{sec:ball-sigma-model} we review the semiclassical ballistic
$\sigma$-model. In sections \ref{sec:universal-limit} and
\ref{sec:pert-equiv-semicl}, repsectively, we explore the field theory approach to
universality, and the connections to periodic orbit theory. A number of issues relating to the derivation of the
theory will be addressed in appendix \ref{sec:regularization}.

\section{Ballistic $\sigma$-model}
\label{sec:ball-sigma-model}

We wish to explore correlations in the level density, $\rho(E)$, of
globally hyperbolic (chaotic) quantum systems, as characterized by the
two point correlation function $R_2(s) \equiv \Delta^2 \langle
(\rho(E+{\Delta s\over 2\pi})- \langle \rho(E+{\Delta s\over
  2\pi})\rangle) (\rho(E)-\langle\rho(E)\rangle) \rangle$. Here,
$\langle\dots\rangle$ denotes averaging over $E$ over an interval
$[E_0-\Eav/2,E_0+\Eav/2]$, where $\Eav \ll E_0$. The goal is to show
that for energies $\omega=s\Delta/\pi \lesssim \hbar \tE^{-1}$ the
function $R_2(s)$ approaches the spectral correlation functions of
RMT. Here, $\tE \equiv \lambda^{-1} \ln(c^2/\hbar)$ is the Ehrenfest
time, $\lambda$ the dominant Lyapunov exponent of the system and $c^2$
some classical reference scale of dimension 'action' whose specific
value will not be of much relevance.~\cite{fn1}

We represent the spectral correlation function in terms of a replica
generating functional (Choosing the replica variant
of the theory is motivated by its high suitability to perturbative
calculations~\cite{smueller2}; it is a matter of a straightforward redefinition of the
field target space to upgrade the formalism to a supersymmetric field
theory.),
\begin{equation}
  \label{eq:11}
  R_2(s) = -{1\over 2}{\rm Re}\,\lim_{R\to 0}{1\over R^2}\partial_s^2 Z(s),
\end{equation}
where $R$ is the number of replicas and the generating function is
defined as 
\begin{widetext}
  \begin{align}
  \label{eq:5}
Z(s) \equiv \int DT\,\exp(-S[T]), \qquad
   S[T]= { \beta\pi \hbar\over 2\Delta} \int_\Gamma (d\bx)  {\rm tr}\Big(T^{-1}\ast \Lambda \{
  H,T\} + {is \Delta\over 2\pi\hbar}\, 
  \Lambda T^{-1} \ast \Lambda T\Big).
\end{align}
\end{widetext}
Eq.~\eqref{eq:5} has the status of an effective field theory in
classical phase space, obtained from a microscopic parent theory --
the energy averaged field integral representations of the microscopic
Green functions~\cite{bsm,ASAA1,ASAA2} -- by elimination of field
configurations fluctuating on short scales $<\hbar$. In
Appendix~\ref{sec:regularization} we will argue that Eq.~\eqref{eq:5}
relates to the microscopic formalism in much the same way the
semiclassical Gutzwiller sum relates to the microscopic Feynman path
integral.

The integration variables in (\ref{eq:5}), $T(\bx) =
\{T^{\alpha\alpha'}(\bx)\}$ are matrix valued fields defined on shells
$\Gamma=\{\bx|H(\bx)=E_0\}$ of constant energy in classical phase
space. Here, $\bx\equiv (\bq,\bp)$ where $\bq$ and $\bp$ are
coordinates and momenta, respectively, $H(\bx)$ is the Hamiltonian
function of the system, and the integral over the energy-shell is
normalized to unity, $\int_\Gamma (d\bx)=1$. For time reversal and
spin rotation invariant systems (orthogonal symmetry class,
$\beta=1$), the 'internal' structure of the matrices
$T^{\alpha\alpha'}$ is described by a composite index
$\alpha=(a,r,t)$, where $a=+/-$ discriminates between the advanced and
retarded sector of the theory, $r=1,\dots,R$ is a replica index, and
$t=1,2$ accounts for the operation of time reversal. Time reversal
symmetry reflects in the relation $\sigma_2^{\rm tr}\ T^T \
\sigma_2^{\rm tr} = T^{-1}$, where $\sigma_i$ are Pauli matrices and
the superscript 'tr' indicates action in time reversal space. For time
reversal non-invariant systems (unitary symmetry, $\beta=2$) no
time-reversal structure exists and $\alpha=(a,r)$. In either case, the
matrices $T$ carry a coset space structure in the sense that
configurations $T$ and $TK$ are identified if $[K,\Lambda]=0$, where
$\Lambda=\sigma_3^{\rm ar}$ and 'ar' stands for action in
advanced/retarded space.

The fluctuation behavior of the fields $T$ in (\ref{eq:5}) is governed
by the classical Liouville operator $\{H,\;\}$ (where $\{\;,\;\}$ is
the Poisson bracket.) Quantum mechanics enters the problem through the
back-door, viz. by the presence of Moyal products '$\ast$' in
(\ref{eq:5}). The Moyal product between two phase space functions $A$
and $B$ is defined as
\begin{equation}
  \label{eq:3}
  (A\ast B)(\bx) = \int {d^f \bx_1\over (\pi \hbar)^f} {d^f \bx_2\over
    (\pi \hbar)^f} \,e^{{2i\over \hbar} \bx_1^T I
  \bx_2} A(\bx+\bx_1) B(\bx+\bx_2),
\end{equation}
where $I \equiv \left(
\begin{smallmatrix}
  0 &\openone\cr -\openone & 0
\end{smallmatrix}
\right)$ is the symplectic unit matrix.  (For all practical purposes,
the definition (\ref{eq:3}) will be more convenient than the standard
representation~\cite{moyal}, $ (A\ast B)(\bx)=\exp({i\hbar\over
  2}\partial^T_{\bx'} I
\partial^\vd_\bx)\; A(\bx')B(\bx)\big|_{x=x'}$.) The presence of the
Moyal product implies that (a) all quantities appearing in the theory
get effectively averaged (smoothened) over Planck cells. Relatedly, (b)
the generator of classical time evolution $\{H,\;\}$ acts on
\textit{distributions} smooth in phase space on scales $\sim \hbar^f$,
rather
than on mathematical points. In the following sections we will
explore how the interplay of hyperbolic dynamics
($\{H,\;\}$) and  Planck cell smearing ($\ast$) determines the output
of the theory.

\section{Universal limit}
\label{sec:universal-limit}

We wish to explore the behavior of the theory (\ref{eq:5}) for
correlation energies $\omega\sim 2\pi \hbar\, t_H^{-1} \ll \hbar\,
\tE^{-1}$ of the order of the inverse Heisenberg time $t_H=2\pi \hbar
\Delta^{-1}$ and much smaller than the inverse of the Ehrenfest
time. For simplicity, we will consider systems with broken time
reversal invariance throughout this section.

For energies $\omega \gg \Delta \Leftrightarrow s\gg 1$, the partition
function $Z(s)$ may be evaluated by perturbative methods.
To prepare the perturbative  expansion of the action, we introduce the
rational parameterization 
\begin{equation}
  \label{eq:12}
  T=\openone + W, \qquad W=
\left(
\begin{matrix}
&-B^\dagger\cr B&
\end{matrix}
\right),
\end{equation}
where the block structure is in advanced/retarded space and the
generators $B$ are $R\times R$ complex matrices. Substitution of this
representation then leads to a series $S[B,B^\dagger]=
\sum_{n=1}^\infty S^{(2n)}[B,B^\dagger]$, where $S^{(n)}$ is of $n$th
order in $B$ and $B^\dagger$. Specifically, 
\begin{equation}
  \label{eq:15}
  S^{(2)}[B,B^\dagger] = t_H \int_\Gamma (d\bx)  \,{\rm tr}(B^\dagger {\cal L}_\omega
  B),  
\end{equation}
where 
\begin{equation}
  \label{eq:13}
  {\cal L}_\omega = -i\omega/\hbar - \{H,\;\}  
\end{equation}
and the Moyal
product stars have been omitted for notational simplicity. Higher
order terms in the action contain traces over (Moyal) products of
matrices $B B^\dagger \dots B^\dagger$. Due to the Planck-cell
'averaging' inherent to the product (\ref{eq:3}), the Wick contraction
of individual matrix elements of $B$ and $B^\dagger$ will generate expressions

\begin{eqnarray}
&&P_\omega(\check{\bx},\check{\bx}') =\int_0^\infty dt\,  e^{i\omega t\hbar^{-1}}
 P_t(\check\bx,\check \bx')
\label{eq:16}
\end{eqnarray}
where 
\begin{eqnarray}
  \label{eq:23a}
 P_\omega \equiv ( -i \omega/\hbar - \{H,\;\})^{-1},\qquad
P_t \equiv e^{t\{H,\;\}},
\end{eqnarray}
are the (retarded) Liouville propagators in the energy and time
representation, respectively, $\bx$ and $\bx'$ are arbitrary points in
phase space and 
$$
f(\check \bx) \equiv {1\over \hbar^f}\int_{\hbar^f } d^f
\by\, f(\bx + \by)
$$ 
is symbolic notation for coordinate averaging over a Planck cell.  We
interpret $ P_t(\,\cdot\,,\check\bx')$ as the dynamical evolution of a
smooth phase space distribution of extension $\hbar^f$ and centered
around $\bx'$.  It has been rigorously shown by methods of symbolic
dynamics~\cite{ruelle} that for times $t< \tE$, these distributions are
centered around the classical trajectory through $\bx'$, i.e. the
dynamics is approximately described by the Liouville evolution of
$\bx'$. However, beyond $t \simeq \tE$, the distribution rapidly (over
time scales comparable to $t_{\rm mix}$) crosses over to a uniform
distribution in phase space. These structures are summarized in the
ansatz
\begin{equation}
  \label{eq:24}
  P_t(\check\bx,\check\bx')\equiv \left\{
    \begin{array}{ll}
      \Omega\,\delta(\bx-\bx'(t)),& t\lesssim \tE,\cr
      1,& t>\tE,
    \end{array}
\right.
\end{equation} 
where $\Omega={h^f\over \Delta}$ is the volume of the energy-shell,
and the normalization is such that for our unit-normalized phase space
integral, $\int_\Gamma (d\bx)\, P_t(\check \bx,\check \bx')=1$.

For energy scales $\omega < \hbar \tE^{-1}$ the dominant contribution
to the time integral in (\ref{eq:16}) comes from large times $\tE<t
\lesssim \hbar \omega^{-1}$. One expects that in this regime phase
space fluctuations will have effectively relaxed to spatial
uniformity.  To describe the damping of inhomogeneous modes in more
explicit terms we follow a prescription formulated by Kravtsov and
Mirlin~\cite{kravtsovmirlin}, then in the context of diffusive
systems: employing the ansatz
\begin{equation}
  \label{eq:17}
  T(\bx) = T_0 (1+W_P(\bx))
\end{equation}
we isolate the
inhomogeneous contents of the fields $T$. Here, $T_0={\rm const.}$
describes the zero mode sector and 
$$
W_P(\bx) \equiv W(\bx) -
W_Q(\bx), \qquad W_Q(\bx) \equiv \int_\Gamma (d\bx) \, W(\bx),
$$     
is a projection onto purely fluctuating field configurations: $\int
(d\bx) \, W_P(\bx)=0$.  Substitution of Eq.~(\ref{eq:17}) into the
action and expansion to second order in $W_P=
\left(\begin{smallmatrix}
  &-B_P^\dagger\\B_P&
\end{smallmatrix}\right)
$ generates the decomposition
\begin{equation}
  \label{eq:18}
  S[T_0,W_P]= S_0[Q_0] + S^{(2)}[B^\vd_P,B_P^\dagger] +
  S_c[T_0,B^\vd_P,B_P^\dagger], 
\end{equation}
where 
\begin{equation}
  \label{eq:19}
  S_0[Q_0]={i\omega\pi \over 2\Delta} {\,\rm tr}(Q_0\Lambda),\qquad Q_0=T_0\Lambda T_0^{-1}
\end{equation}
is the zero mode action, and 
\begin{equation}
  \label{eq:20}
  S^{(2)}[B^\vd_P,B_P^\dagger]=  t_H \int_\Gamma
  (d\bx)  \,{\rm tr}(B_P^\dagger {\cal L}_0 B^\vd_P),
\end{equation}
the quadratic action of the fluctuating fields governed by the
generator  ${\cal L}_0 = {\cal L}_{\omega=0} = \delta-\{H,\;\}$,
$\delta\searrow 0$ of the time integrated dynamics. Finally, 
\begin{eqnarray}
  \label{eq:21}
&&   S_c[T_0,B^\vd_P,B_P^\dagger]= -{i \omega t_H\over 2\hbar} \int_\Gamma (d\bx) \times\\
&&\times {\,\rm
     tr}\left(( T_0^{-1} \Lambda T_0)^{11}B^\dagger_P B_P^\vd  - (
   T_0^{-1} \Lambda T_0)^{22}B^\vd_PB_P^\dagger\right)(\bx),\nonumber  
\end{eqnarray}
where the superscripts refer to advanced-retarded space. Following
Ref.~\onlinecite{kravtsovmirlin}, we expand in $S_c$, retaining only
contributions of minimal order in $\omega$:
\begin{eqnarray*}
&&  S_0[Q_0]\to S_{\rm eff}[Q_0]= S_0[Q_0] +  \\
&&\hspace{.3cm}+  \langle S_c[
  T_0,B^\vd_P,B_P^\dagger] \rangle- {1\over 2} \langle (S_c[
  T_0,B^\vd_P,B_P^\dagger])^2 \rangle,
\end{eqnarray*}
where $\langle \dots \rangle \equiv \int D(B_p^\vd,B_P^\dagger) \,
\exp(- S^{(2)}[B_P^\vd,B_P^\dagger])\,(\dots)$. While the contraction of the
$B$'s in the first contribution to the second line vanishes in the
replica limit, the second
term generates the effective action
\begin{eqnarray}
  \label{eq:22}
&& S_{\rm eff}[Q_0]= S_0[Q_0] -{\omega^2\over 16\hbar^2}({\rm
  tr}(Q_0\Lambda))^2\times \nonumber\\ 
&&\hspace{.5cm}\times 
\int_\Gamma (d\bx) (d\bx') (P_P)_0(\bx, \bx')(P_P)_0( \bx',
\bx),
\end{eqnarray}
where, again, the subscript $P$ stands for projection onto the fluctuating sector:
\begin{eqnarray}
\label{eq:aa}
&&(P_P)_0(\bx,\bx')\equiv \Big(1-\int_\Gamma {(d\bx)}\Big)
\Big(1-\int_\Gamma {(d\bx')}\Big)P_0(\bx, \bx')\nonumber\\  
&&\hspace{.5cm}= \int_0^{\infty} dt
\,\left(\delta(\bx-\bx'(t)) - \Omega^{-1}\right),
\end{eqnarray}
and in the last line we have switched to a time representation. Notice
that the Liouville propagator in Eq.~(\ref{eq:22}) is evaluated
on single phase space points, $\bx$ and $\bx'$, rather than on
Planck-cells $\check\bx$ and $\check\bx'$. This is because the phase
space integral of the Moyal-product of two operators collapses to the
ordinary product~\cite{jmueller}, i.e.
\begin{align*}
\int_\Gamma (d\bx) {\,\rm tr}&\left(( T_0^{-1} \Lambda
  T_0)^{11}B^\dagger_P \ast B_P^\vd\right)(\bx) \\
=&
\int_\Gamma (d\bx) {\,\rm tr}\left(( T_0^{-1} \Lambda T_0)^{11}B^\dagger_P(\bx) B_P^\vd(\bx)\right). 
\end{align*} 
Substituting the result Eq.~(\ref{eq:aa}) into Eq.~(\ref{eq:22}),
we obtain
\begin{eqnarray}
  \label{eq:23}
&& S_{\rm eff}[Q_0]= S_0[Q_0] -{\tE\Omega\over 16\hbar^2}(\omega{\rm
  tr}(Q_0\Lambda))^2\times \nonumber\\ 
&&\hspace{.5cm}\times 
 \int_0^{\infty} dt \int_\Gamma (d\bx)\,  \left(\delta(\bx-\bx(t))-\Omega^{-1}\right).
\end{eqnarray}
According to Eq.~\eqref{eq:23}, the correction term is given  
by the integrated weight of 
periodic orbits $\bx\stackrel{t}{\to}\bx$ of duration $t$, minus the
total phase volume. Using that~\cite{gaspard}
$$
\lim_{T\rightarrow+\infty} \int_0^{T} dt\, \int_\Gamma (d\bx)\,  \Omega\, \delta(\bx-\bx(t)) = T,
$$ 
one concludes that this term cancels, i.e. to lowest order in the
expansion, fluctuations of the inhomogeneous modes
do not change the universal zero-mode action. At higher orders in the
expansion, we are met with expressions (cf. Eq.~\eqref{eq:16})
\begin{eqnarray*}
&&  (P_P)_\omega(\check\bx,\check \bx') =\\
&&\hspace{.2cm}=\int_0^\infty dt\,  e^{i{\omega t\over \hbar}}
  \Big(1-\int_\Gamma {(d\bx)}\Big) \Big(1-\int_\Gamma
  {(d\bx')}\Big)P_t(\check\bx,\check \bx')\simeq\\
&&\hspace{.2cm}\simeq\int_0^{t_E} dt\,  e^{i{\omega t\over \hbar}}
  \Big(\Omega\delta(\bx-\bx'(t))-1),
\end{eqnarray*}
where Eq.~\eqref{eq:24} was used. The important point here is the
truncation of the time integral at $t_E$. This means that at $n$th
order in the expansion in fluctuation propagators, $P_P$, corrections
$\sim (\omega t_E/\hbar)^n$ will be generated. Our analysis above
shows that terms with $n< 3$ cancel, i.e. corrections to the zero mode
action can arise only at $\mathcal{O}(\omega t_E\hbar^{-1}) \sim
\mathcal{O}(t_E/t_H)^3$,
$$
S_{\rm eff}[Q_0]=S_0[Q_0] + \mathcal{O}(\omega t_E/\hbar)^{n\ge 3}.
$$
This observation is consistent with the periodic orbit analysis of
Ref.~\cite{a1} where it has been shown that the leading correction to
the universal result scales as $(\omega t_E/\hbar)^3$.

Summarizing, our analysis shows that at correlation energies $\omega <
\hbar/\tE$ inhomogeneous fluctuations get effectively damped out and
the field theory collapses to an integral over the zero mode. This
means that RMT results (plus weak corrections in the parameter
$(\omega \tE/\hbar)^{n\ge 3}$) will be obtained for spectral
correlation functions, and other observables probing the long time
behavior of the system. Our 'mode damping' approach to universality is
complementary to semiclassical analysis~\cite{aa+haake}, where the
spectral correlation functions are constructed explicitly, by
summation over infinite hierarchies of periodic orbits. The connection
between these approaches will be explored in the next section.

\section{Perturbative equivalence to semiclassics}
\label{sec:pert-equiv-semicl}

Having shown that the ballistic $\sigma$-model crosses over to the
universal zero dimensional model at low energies, we now approach
the problem from a  different perspective. We will perform a
straightforward perturbative expansion around the high energy saddle
point  $\Lambda$ to elaborate
on parallels between the field theory (\ref{eq:5}) and the periodic
orbit approach to spectral correlations. Focusing on the lowest
order quantum corrections to the spectral form factor (the so-called
Sieber-Richter term~\cite{sieber_richter01}), we will show that the two formalism are structurally similar, to an
extent that the correlations of individual periodic orbits can be
reproduced from the field theory formalism.

Throughout this section, we will consider time reversal invariant
systems, i.e. $\beta=1$. 

\subsection{Semiclassics}
\label{sec:semiclassics}

For the benefit of non-expert readers, we begin with a brief review of
recent semiclassical results. Consider the Fourier transform of
the spectral correlation function, 
\begin{equation}
  \label{eq:6}
  K(\tau) \equiv {1\over \pi} \int ds \, R_2(s) e^{-2i s \tau}.
\end{equation}
The RMT result for the $\beta=1$ form factor reads as
\begin{eqnarray}
K(\tau) = \left\{
  \begin{array}{ll}
    2 \tau - \tau \ln(1+2\tau),&\tau<1\cr
2-\tau\ln\left({2\tau+1\over 2\tau-1}\right),& \tau\ge 1. 
  \end{array}\right.\label{eq:10}  
\end{eqnarray}
Specifically, the short time expansion of the spectral form factor
in $\tau\ll 1$ (corresponding to the expansion of the spectral
correlation function in $\omega\gg \Delta$) starts as
\begin{equation}
  \label{eq:8}
 K(\tau ) = 2\tau - 2 \tau^2 + \dots.  
\end{equation}
In semiclassics one aims to reconstruct that expansion from the
Gutzwiller double sum over periodic orbit pairs $(\alpha,\alpha')$,
\begin{equation}
  \label{eq:7}
  K(\tau) = \left\langle \sum_{\alpha\alpha'} A_\alpha
    A^\ast_{\alpha'} e^{{i\over
        \hbar}(S_\alpha-S_{\alpha'})} \delta\left(\tau -
      {t_\alpha+t_{\alpha'}\over 2t_H}\right)\right\rangle,
\end{equation}
where $\langle \dots \rangle$ is an
average over orbit energies (and a small interval of orbit times.)
and $A_\alpha, S_\alpha$, and $t_\alpha$ are the stability amplitude,
the action, and the time of traversal of orbit $\alpha$, respectively.

The first term in (\ref{eq:8}) obtains from Berry's diagonal
approximation,
$$
K^{(1)}(\tau)\equiv 2\tau,
$$
i.e. an approximation that retains only identical orbits
$\alpha=\alpha'$ and mutually time reversed
$\alpha=\overline{\alpha}'$ orbits, and uses the Hannay-Ozorio de
Almeida~\cite{HOsum} sum rule $\big\langle \sum_\alpha
|A_\alpha|^2\delta(\tau -t_\alpha/t_H)\big \rangle=\tau$ to determine
the weight of the remaining orbit sum.

\begin{figure}[htbp]
   \centering
   \resizebox{3in}{!}{\includegraphics{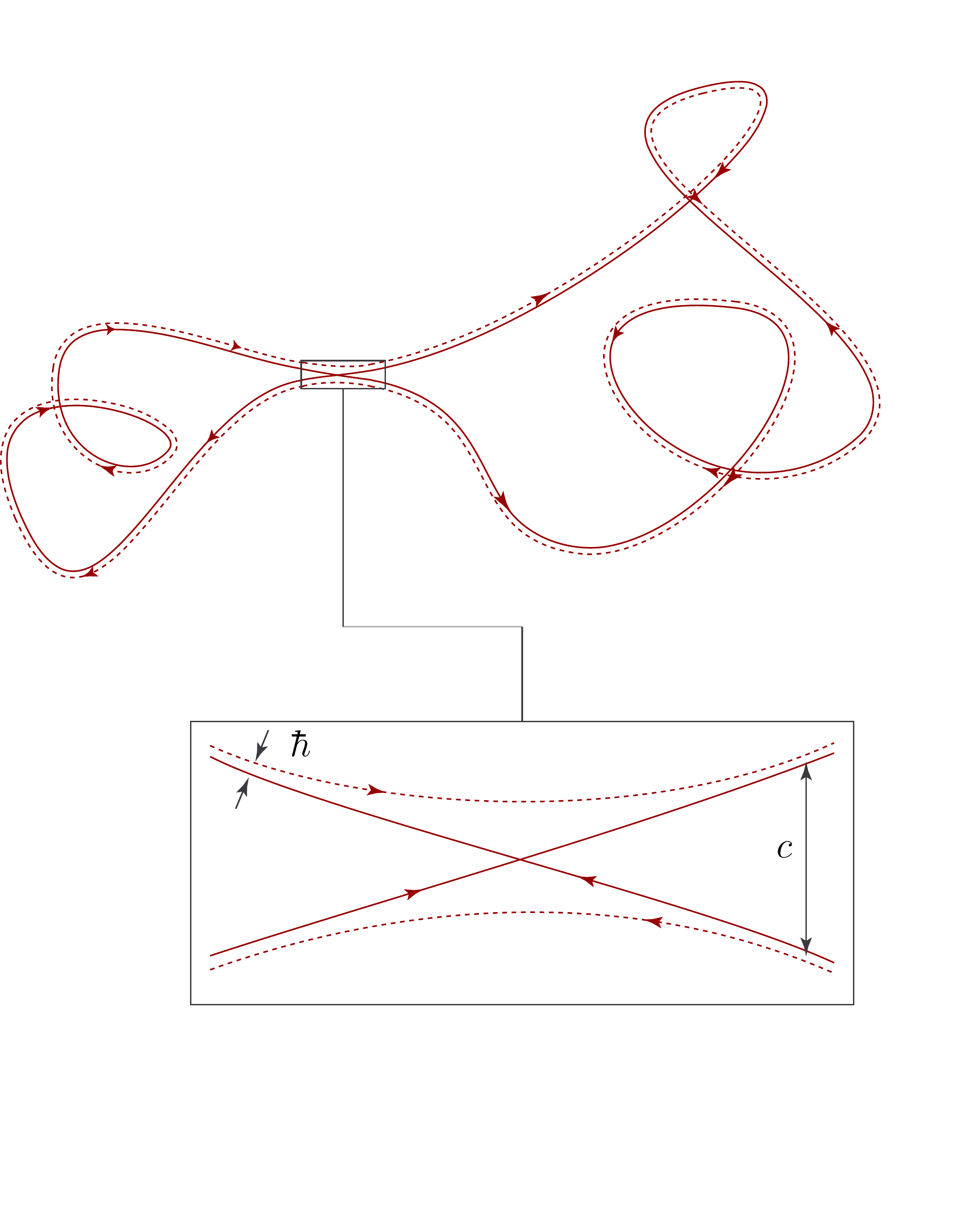}}
   \caption{Cartoon of an orbit pair contributing to the first quantum
   correction to the spectral form factor.}
   \label{sieber_richter}
 \end{figure}

 The structure of the leading order (in $\tau$) quantum corrections to
 the spectral form factor was first identified by Aleiner and
 Larkin~\cite{aleinerlarkin}: two orbits differing in the presence or
 absence of a small angle self intersection in configuration space
 interfere to provide a stable contribution to the double sum
 (cf. Fig.~\ref{sieber_richter}.) For the sake of later comparison to
 the $\sigma$-model formalism, we briefly review the quantitative
 computation of this contribution~\cite{sieber_richter01} in the
 invariant language of hyperbolic phase space coordinates~\cite{sc2}
 for a system with $f=2$ degrees of freedom.

\begin{figure}[htbp]
   \centering
   \resizebox{3.5in}{!}{\includegraphics{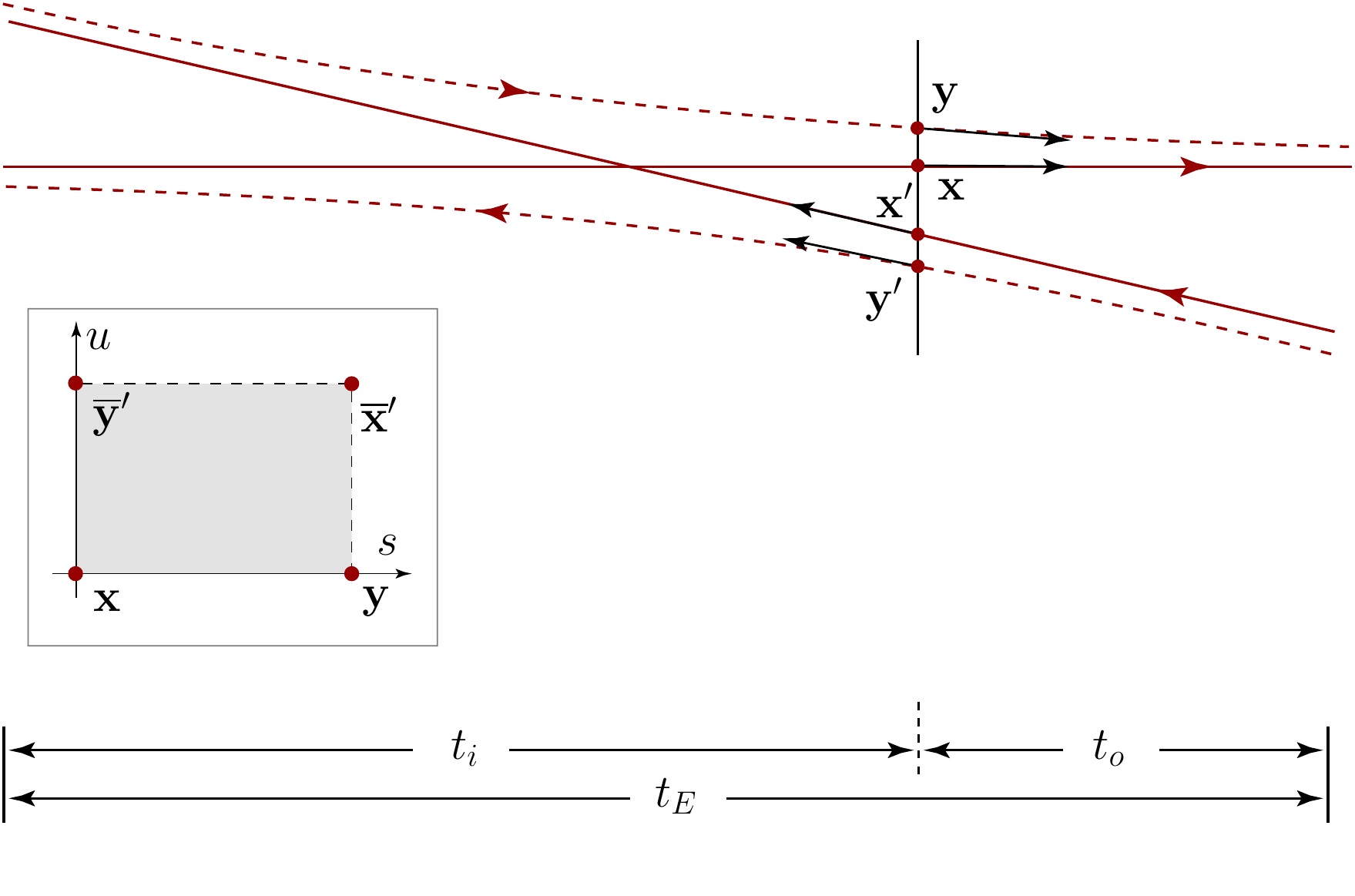}}
   \caption{Phase space representation of the encounter
     region. Discussion, see text.}
   \label{phase_space}
 \end{figure}

 Imagine a Poincar\'e section cutting through the encounter region
 where the participating orbit segments have their (avoided) self
 intersection (cf. Fig.~\ref{phase_space}.) Choosing one of the
 piercing points, $\bx$, as the origin of a local coordinate system
 (cf. inset of Fig.~\ref{phase_space}), we introduce $2f-1=3$ energy
 shell coordinates $(t,u,s)$, where $t$ is a time-like coordinate
 running along the trajectory through $\bx$, and $u/s$ are coordinates
 along the locally unstable/stable directions (the existence of the
 latter guaranteed by the assumption of global hyperbolicity.) One may
 then show~\cite{sieber_richter01,sc2} that the contribution of all orbit
 pairs containing a single encounter region is given by
\begin{equation}
  \label{eq:9}
 K^{(2)}(\tau)=2 {\tau^2\over h}  \int du ds \,e^{i\hbar^{-1} us} {t-2t_{\rm
    enc}(u,s)\over 2 t_{\rm enc}(u,s)} = -2\tau^2.
\end{equation}
Here, $t_{\rm enc}(u,s) = \lambda^{-1} \ln(c^2/us)$ is the time of
traversal through the encounter region, $\lambda$ the (self-averaging)
Lyapunov exponent of the system, and $c^2$ some classical action
scale. We note that (i) the dominant contributions to the integral are
of order $us \sim \hbar$ implying that $t_{\rm enc}(u,s) \sim
\lambda^{-1} \ln(c^2/\hbar)$ is of the order of the Ehrenfest time,
(ii) the product $us$ is an invariant of the motion (Liouville's
theorem), which means that the choice of the Poincar\'e section
entering the construction is arbitrary, and (iii) the only
surviving contribution to the integrand in (\ref{eq:9}) are of order
${\cal O}(t_{\rm enc}(u,s)^0)$. (Terms containing $t_{\rm enc}$,
either from the numerator, or the expansion of the denominator can be
shown to effectively oscillate to zero in the semiclassical limit~\cite{smueller1,smueller2,aa+haake}.)

In later work, the perturbative
analysis of action correlations has been extended first to
cubic and then to infinite order~\cite{smueller1,smueller2,smueller3} in the
$\tau$-expansion. Employing an unconventional representation of the
two-point correlation function~\cite{aa+haake} it has also been
possible to reconstruct the bottom part of Eq.~(\ref{eq:10}), which is
not accessible in terms of a straightforward series expansion of the
conventional two-Green function representation.

\subsection{Field Theory}
\label{sec:field-theory}

Eq.~(\ref{eq:9}) encapsulates the phase volume available to individual
pairs of interfering trajectories (pre-exponent) and the corresponding
action correlation (exponent). Extending the analysis of
Ref.~\onlinecite{jmueller}, we will show below that the same
information is stored in the effective action (\ref{eq:5}).  Although
our discussion will be restricted to first order in perturbation
theory in $\tau$ (the extension of the analysis to higher orders in
perturbation theory is left for future work), it suggests a general equivalence between  periodic
orbit theory  and the perturbative expansion of
the $\sigma$-model.

In the time reversal invariant case considered here, the blocks
$B$ in (\ref{eq:12}) are 
$(2R\times 2R)$-matrices subject to the  time
reversal symmetry condition $B^\dagger= -\sigma_2^{\rm tr}\ B^T
\sigma_2^{\rm tr}$.

Substitution of this representation into the action (\ref{eq:5})
generates the first order correction to the spectral correlation
function 
$$
R_2(s)^{(2)} = - {\rm Re}\, \lim_{R\to 0}{t_H\over (2R)^2} \partial_s^2 
\int_\Gamma (d\bx) \left\langle \,{\rm tr}(B^\dagger B B^\dagger {\cal
    L}_\omega B)\right\rangle ,
$$
where we have omitted the Moyal product stars for the sake of
notational simplicity, and averaging is over the quadratic action,
$
\langle \dots \rangle \equiv \int {\cal D}(B^\dagger,B)\,
e^{-S^{(2)}[B,B^\dagger]}(\dots)
$. 
Doing the Gaussian integrals over matrix elements of $B$, we obtain
(cf. Ref.~\onlinecite{jmueller} for technical details),
\begin{widetext}
\begin{eqnarray*}
  R^{(2)}(s) =  {\rm Re}\, {\Omega^2\over t_H} \partial_s^2
\int_\Gamma (d\bx) \int {d\by_1d\by_2\over (2\pi\hbar)^{2(f-1)}}\,
e^{{i\over\hbar} \by_2^T I \by_1}
P_\omega(\bx+\by_1/2,\overline{\bx-\by_1/2}){\cal
  L}_{\omega,\bx+\by_2/2}
P_\omega(\overline{\bx+\by_2/2},\bx-\by_2/2).
\end{eqnarray*}
Here, $\overline{\bx}\equiv (\bq,-\bp)^T$ is the time reversed of the
phase space point $\bx=(\bq,\bp)^T$, the integrals over $\by_1$ and
$\by_2$ represent the generalization of the Moyal product (\ref{eq:3})
to the trace of a product of four operators and the subscript $\bx$ in
${\cal L}_{\omega,\bx}$ indicates on which coordinate of the
propagator $P_\omega(\bx,\bx')$ the derivatives
of the Liouville operator act. To make further progress, we Fourier
transform the expression above whereupon it transmutes to the first
quantum correction to the spectral form factor,
\begin{eqnarray}
K^{(2)}(\tau) =-2\tau^2\Omega^2 
 \int_\Gamma (d\bx) \int {d\by_1 d\by_2e^{\frac{i}{\hbar}\by_2^T I \by_1 }\over(2\pi\hbar)^{2(f-1)}}\,
\int_0^{t} dt' 
P_{t-t'}(\bx+\by_1/2,\overline{\bx- \by_1/2})
{\cal L}_{t',\bx+\by_2/2} P_{t'}(\overline{\bx+\by_2/2},\bx-\by_2/2),  
\label{eq:14b}
\end{eqnarray}

\end{widetext}
Here, ${\cal L}_t\equiv \partial_t - \{H,\;\}$ and $P_t$ is the
propagator in time representation (cf. Eq.~\eqref{eq:23a}.) We proceed by
introducing a coordinate system that has $\bx$ as its origin and $\bx
+ \bx'\leftrightarrow (r,u,s)$, where $r$ is the coordinate of $\bx'$
along the classical trajectory running through $\bx$ and $u$ and $s$
parameterize the components of $\bx'$ in the locally stable and
unstable direction around $\bx$ (see Fig.~\ref{phase_space}.)
We 
assume~\cite{fn2} that the
function $P_t(\overline{\bx+\by/2},\bx-\by/2)$ exhibits the following
characteristics: (i) $P_t(\overline{\bx+\by/2},\bx-\by/2)=P_t(u)$
depends only on the unstable component of $\by$. Indeed, this
component controls the rate at which the trajectory starting at
$\bx-\by/2$ deviates from that through $\bx$. This deviating component
(rather than the approaching component, $s$) determines the spatial
and temporal structure of the Lyapunov region. Conversely,
$P_t(\bx+\by/2,\overline{\bx- \by/2})$ depends only on $s$. (ii) In
order for $P_t(\overline{\bx+\by/2},\bx-\by/2)$ to become
non-vanishing, the two trajectories through $\bx-\by/2$ and
$\overline{\bx+\by/2}$ must have left the Lyapunov region around $\bx$
(cf. the 'large scale picture' Fig.~\ref{sieber_richter}). This takes
a time of order $2\times t(u)$, where
$$
 t(u)\equiv 
\lambda^{-1}\ln(c/u)
$$ 
accounts for 'half' of the encounter time. Additional to this time,
some time of classical duration, roughly of order $t_{\rm mix}$, is
required to 'tie' the outgoing and incoming trajectory segments to a
closed link. These assumptions are summarized in the ansatz
\begin{equation}
  \label{eq:14}
  P_t(\overline{\bx+\by/2},\bx-\by/2)=\Omega^{-1}\tilde
  \Theta(t-2 t(u)),
\end{equation}
where $\tilde \Theta$ is a smeared step function interpolating between
zero and unity over a time interval of order $t_{\rm mix}$. Neither
the detailed structure of $\tilde \Theta$, nor the exact value of
$t(u)$ are of further relevance. We note, however, that the
postulated independence of $P_t$ of the 'longitudinal' coordinate,
$r$, implies stationarity of the long time probability distribution
$P_{t>t(u)}$ under the Hamiltonian flow, $\{H,P_{t>t(u)}\}=0$.

Using that ${\cal L}_t P_t \simeq \partial_t P_t$, the first quantum
correction becomes ($f=2$)
\begin{widetext}
\begin{align}
K^{(2)}(\tau) &=- {2\tau^2\Omega^2\over(2\pi\hbar)^{2}}
 \int_0^{t} dt' \int du_1 ds_1 du_2 ds_2 \,
e^{\frac{i}{\hbar}(u_1 s_2-s_1 u_2)}
P_{t-t'}(s_1)
\partial_{t'} P_{t'}(u_2)\nonumber\\
&=
{\tau^2\Omega^2\over 2\pi\hbar}
 \int_0^{t} dt' \int du ds \,
e^{\frac{i}{\hbar}u s}
P_{t-t'}(s)
\partial_{t(u)} P_{t'}(u)\nonumber\\
&\simeq
  {\tau^2\over 2\pi\hbar}
\int du ds \,
e^{\frac{i}{\hbar}u s}
\partial_{t(u)} (t- 2t(u) - 2t(s))\nonumber\\
&=
  {2\tau^2\over 2\pi\hbar}
\int du ds \,
e^{\frac{i}{\hbar}u s}
\partial_{t_{\rm enc}(u,s)} {t- 2t_{\rm enc}(u,s)\over 2}\nonumber\\
&=
 {2\tau^2\over 2\pi \hbar}
 \int du ds \,
e^{\frac{i}{\hbar}u s}
{t- 2t_{\rm enc}(u,s)\over 2t_{\rm enc}(u,s)},\label{eq:14a}
\end{align}
\end{widetext}
in agreement with the semiclassically derived expression
(\ref{eq:9}). In the first line we used that, under the conditions
stated above, the long time probability distribution is invariant
under the action of the Liouville operator. In the fourth line, we
introduced the full encounter time, $t_{\rm enc}=t(u)+t(s)$, and the
last equality is based on the above mentioned vanishing of any power
$(t_{\rm enc}(u,s))^n$ upon integration against the oscillatory kernel
$\sim\exp(i\hbar^{-1} us)$.

\begin{figure}[htbp]
   \centering
   \resizebox{3.5in}{!}{\includegraphics{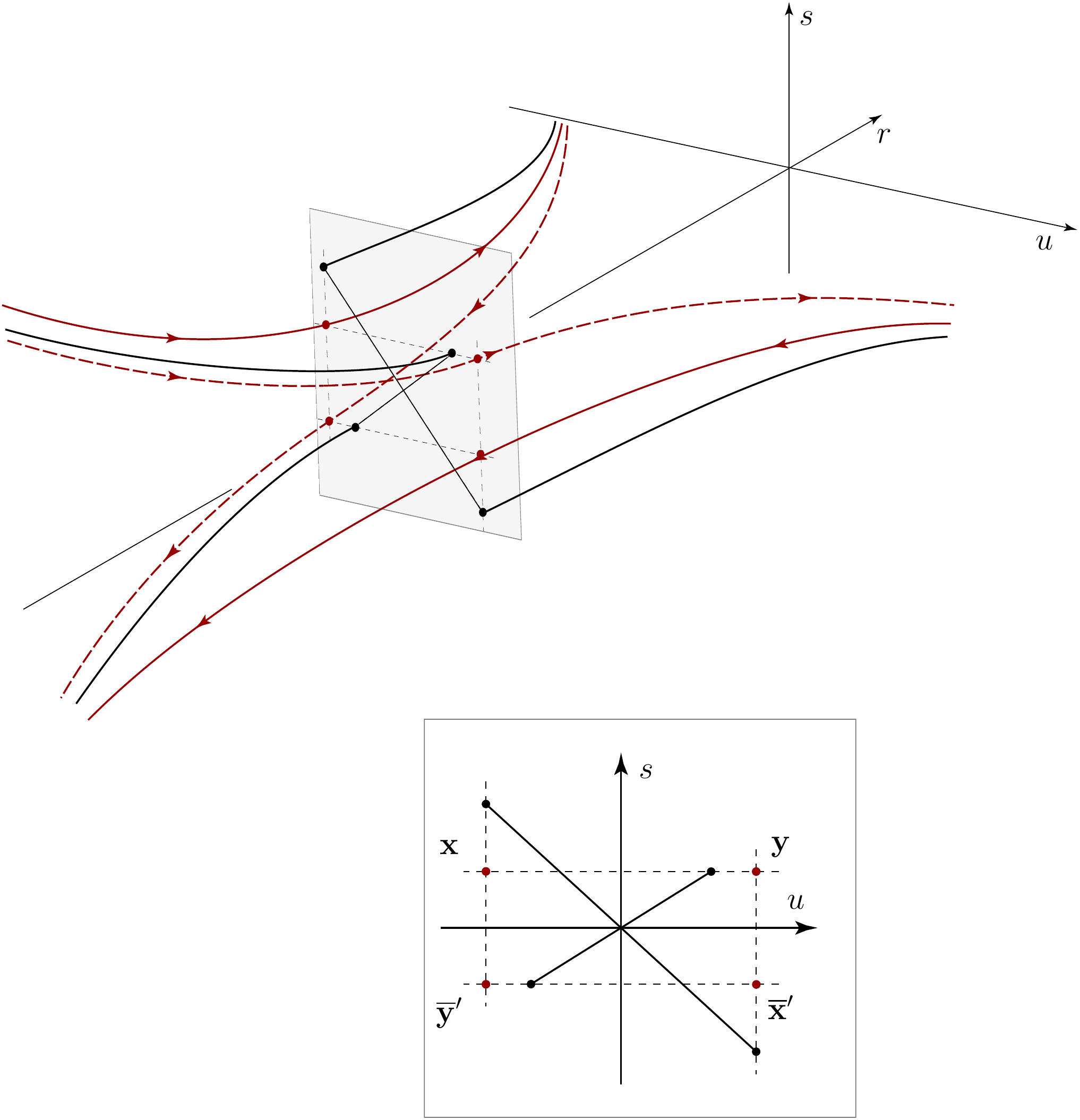}}
   \caption{Assignment of propagator coordinates (lines terminating in
   thick points) to Feynman amplitudes (lines continued through the
   Poincar\'e section).}
   \label{prop_traj}
 \end{figure}

 The derivation above demonstrates that field theory and semiclassics,
 resp., allocate the same phase space volume to single encounter
 processes. We finally show that the field theoretical
 expression~(\ref{eq:14a}) indeed affords an interpretation in terms
 of individual periodic orbits.  To this end, we consider a Poincar\'e
 section through the encounter, as shown schematically in
 Fig.~\ref{prop_traj}. The four lines terminating in the two pairs of
 connected points in the phase space plane represent segments of
 classical trajectories beginning at $\bx+\by_{1,2}/2$ and
 $\overline{\bx-\by_{1,2}/2}$, resp., i.e. idealized classical
 trajectories corresponding to the arguments of the field theory
 propagators in~(\ref{eq:14b}). Each of these points has a stable
 ($s$) and an unstable ($u$) coordinate. The fact that the propagators
 $P$ are retarded implies that we can distinguish between 'incoming'
 and 'outgoing' terminal points.  Now, consider the behavior of a
 trajectory carrying the stable coordinate of an incoming terminal
 propagator point, and the unstable coordinate of a terminal point of
 the other propagator pair (cf. inset of Fig.~\ref{prop_traj}). The
 trajectory running through this reference point will interpolate
 between the propagator stretches involved, and is naturally
 interpreted as part of a closed loop. The analogous definition of
 three more points (see inset of Fig.~\ref{prop_traj}) plus associated
 trajectory stretches leads to the identification of a periodic orbit
 and its topologically distinct partner orbit. The action difference
 between these two orbits is but the product $us$. This construction
 implies an interpretation of the field theory expression
 (\ref{eq:14a}) purely in terms of periodic orbits. The correspondence between the two formalisms is
 established \textit{before} integration over phase space coordinates
 $(u,s,t)$: apparently, the action (\ref{eq:5}) encapsulates
 detailed information about the hyperbolic dynamics and action
 correlations of individual Feynman amplitudes.

\section{Summary and Discussion}
\label{sec:summary-discussion}

In this paper we have considered the semiclassical ballistic
$\sigma$-model as an effective theory of quantum chaos. Defined in
classical phase space, the semiclassical $\sigma$-model is a field
theory whose dynamics is driven by the classical Liouville
operator. Quantum mechanics enters through  a structure known as
non-commutativity. In practice, non-commutativity (i) limits the
maximal resolution of the theory to structures of the order of the
Planck cell, and (ii) leads to the appearance of characteristic
phases, which play a role analogous to the action correlations of
periodic orbit theory.

We have shown in perturbation theory that the $\sigma$-model describes
spectral correlations in far reaching analogy to semiclassics. The
fact that semiclassics and field theory attribute the same weight to
{\it individual} orbit correlations in phase space
(cf. Eqs.~(\ref{eq:9}) and (\ref{eq:14a})) makes one suspect that the
$\sigma$-model fully encapsulates the information carried by the
Gutzwiller double sum. However, this expectation has not been proven
beyond first order in perturbation theory.

Perhaps most importantly, we have shown that the semiclassical
$\sigma$-model provides for an efficient description of
the crossover to the universal regime of RMT correlations at energies
below the inverse of the Ehrenfest time. Adapting a technique
originally developed by Kravtsov and Mirlin to describe spectral
correlations in disordered systems, we saw that in chaotic
systems---in contrast to disordered systems---quadratic inhomogeneous
fluctuations do not give relevant corrections to the universal
action. Extending this approach to higher orders in the inhomogeneous
modes, a systematic quantitative estimate for the corrections to RMT
spectral statistics in individual chaotic systems can, in principle, be
obtained. Methodologically, this way of approaching the universality
phenomenon is complementary to (and arguably more economical than) the
infinite order summations over correlated orbit pairs of
Refs.~\onlinecite{smueller1,smueller2}.

We have enjoyed fruitful discussions with P. Brouwer and F. Haake and
thank S. Heusler for critical reading of the manuscript. Work
supported by the Sonderforschungsbereich SFB/TR12
of the Deutsche Forschungsgemeinschaft.\\

\appendix

\section{Regularization}
\label{sec:regularization}
The $\sigma$-model discussed in the foregoing sections has the status
of an {\it effective} theory, obtained from an underlying 'bare'
theory by elimination of rapid fluctuations. In this section we
discuss the status of that reduction. Comparing to semiclassics, we
will argue that the projection onto an effective field theory
resembles the passage from the Feynman path integral to the
semiclassical Gutzwiller sum.

The action of the bare
ballistic $\sigma$-model~\cite{bsm,ASAA1,ASAA2} is given by
\begin{equation}
  \label{eq:1}
  S_{\rm qu}[T]= -{i\beta\pi\over 2\Eav}{\rm tr}\,\left(T^{-1} \Lambda [\hat H,T] -  
{\Delta s\over 2\pi}\,
  \Lambda T^{-1}\Lambda T\right),
\end{equation}
where, $\hat H$ is the Hamiltonian operator of the system and ${\rm
  tr}$ is a trace over Hilbert space, projected to an energy strip of
with $\Eav$ centered around $E_0$. The integration variables
$T=\{T_{\bq \bq'}^{\alpha\alpha'}\}$ are operators in a product
Hilbert space spanned by real space coordinates, $\bq$, and 'internal'
coordinates, $\alpha$. (The internal structure of the matrices $T_{\bq
  \bq'}$ has been discussed in section \ref{sec:ball-sigma-model}
above.)

The action $S_{\rm qu}[T]$ is obtained from the rigorous functional
integral representation of the two level correlation function after
(a) a saddle point approximation (which generates the nonlinear
constraint $Q^2=\openone$, where $Q=T \Lambda T^{-1}$) and (b) first
order expansion in commutators $[\hat H,T^{-1}]$. As we will argue in
section \ref{sec:zero-modes} below, these two approximations are
largely immaterial.

\subsection{Semiclassical Model}
\label{sec:semiclassical-model}

Preliminary contact with the (semi)classical contents of the theory is
made by switching to a Wigner representation. Upon Wigner
transformation of Hilbert space operators $A_{\bq_1,\bq_2} \to
A(\bx)\equiv \int d\Delta \bq\,e^{i\hbar^{-1} \Delta\bq\cdot \bp}
A_{\bq+\Delta\bq/2,\bq - \Delta\bq/2}$, the operators $T_{\bq,\bq'}
\to T(\bx)$ transmute to the fields in classical phase space discussed
above.~\cite{fn3} Further, ${\rm tr}(\dots)
\to \int_{\Gamma_{\Eav}} d^f \bx\, (\dots)$ becomes an integral over a
phase space energy shell $\Gamma_{\Eav}$, centered around $E_0$ and of
thickness $\Eav$. We are thus led to the phase space representation
\cite{jmueller}
\begin{widetext}
\begin{equation}
  \label{eq:2}
  S_{\rm ps}[T]= -{i\beta\pi\over 2\Eav} \int_{\Gamma_{\Eav}}{d^f\bx\over h^f} \, {\rm tr} \left(T^{-1}\ast \Lambda
  [H\stackrel{\ast}{,}T] - {\omega\over 2}\, 
  \Lambda T^{-1} \ast \Lambda T\right),
\end{equation}
\end{widetext}
where the asterisks stand for Moyal products, as usual.

Importantly, the field theory (\ref{eq:2}) does not contain mechanisms
inhibiting the buildup of rapid fluctuations.  The action $S[T]$ of
fields fluctuating on quantum scales $\sim \hbar^{\alpha}$, $\alpha\ge
1$ in classical phase space is qualitatively different from the
semiclassical action considered above: for fields fluctuating
transverse to the energy layer of thickness $E_{\rm av} \sim
\hbar/t_0\sim \hbar$ reduction of the phase space integral to an
integral over a single energy 'shell' is no longer possible. Further,
the series expansion of the Moyal product
$$
 [H\stackrel{\ast}{,}T]= i\hbar \{H,T\} + \hbar^3\, {\cal O}(\partial^3
 H \partial^3 T). 
$$
shows that for fields fluctuating on scales $\sim \hbar^{\alpha}$,
$\alpha\ge 1$, the approximation of the quantum commutator by the
Poisson bracket is no longer permissible.

A save way to eliminate those rapid field fluctuations is by
adding a weak random potential to the Hamiltonian. An ensemble average
over that randomness will generate a second order differential
operator which may be tailored so as to effectively remove fast field
fluctuations. As discussed in Ref.~\onlinecite{dis} a 'quantum random
potential' $V_q\sim \hbar$, vanishing in the classical limit (and
parametrically in $\hbar$ weaker 
weaker than the 'regulator' suggested by Aleiner and
Larkin~\cite{aleinerlarkin,aleinerlarkin2}) suffices
to remove field fluctuations on phase space scales $\sim \hbar$.

Less rigorously, one may argue that the action $S$ of rapidly
fluctuating fields will lead to highly oscillatory exponents
$\exp(iS)$ which likely tend to average out: a glance at
Eq.~(\ref{eq:5}) shows that the largest contributions to the action of
the problem, set by the largest value of the correlation energy
$\omega\sim \hbar/t_{\rm mix}$, scale as $ \Delta^{-1}\times
\hbar$. However, fields fluctuating on scales $\sim \hbar^{\alpha}$,
$\alpha\ge 1$ will lead to contributions of $\Delta^{-1}\times \hbar^{n(1-\alpha)}$, where the odd index $n$ designates the
order of the Moyal expansion. One may argue that in the semiclassical
limit, these terms generate strong phase cancellations which render
the contribution of strongly fluctuating fields effectively
meaningless. (For a caveat in this argumentation, see
section \ref{sec:zero-modes} below.)

We note that the above phase cancellation argument resembles the
logics inherent to the stationary phase derivation of the Gutzwiller
trace formula from the Feynman path integral; there, too, avoidance of
rapid fluctuations is the dominant principle.  More specifically,
Gutzwiller's trace formula is based on a stationary phase
approximation to the Feynman path integral in $S/\hbar$ (where $S$
stands for the typical value of a classical action.) This
approximation effectively averages over fine structures on scales
$\hbar^\alpha, \alpha\ge1$. Relatedly, the semiclassical analysis of
spectral correlations will be oblivious to the averaging of the system
Hamiltonian over 'quantum' perturbations $V_q\sim \hbar$. (Not
affecting the classical dynamics, such perturbations merely change the
phases weighing individual periodic orbits. In the evaluation of the
Gutzwiller double sum of spectral correlations, these phases cancel
out.~\cite{fn4}) 

Summarizing, the semiclassical $\sigma$-model \eqref{eq:5} considered
in the main text obtains as a projection of the 'quantum'
$\sigma$-model \eqref{eq:1} onto the sector of fields fluctuating on
scales $> \hbar$. This projection may be effected either by averaging
the system over a quantum random potential of strength $\sim \hbar$,
or by alluding to the prospected irrelevance of strongly fluctuating
actions in the semiclassical limit (i.e. ad hoc restriction of the
functional integral to fields fluctuating on scales $>\hbar$.)

One thus reduces the microscopic theory (equivalent to the full
Feynman path integral representation of Green function correlators) to
an effective theory (likely equivalent to the Gutzwiller approximation
of the Feynman path integral.)

\subsection{Zero modes}
\label{sec:zero-modes}

While the above phase cancellation arguments apply to 'generic'
field configurations, there is one family of rapidly fluctuating
configurations that deserves separate consideration: the quantum
action (\ref{eq:1}) is nullified by a large number of \textit{zero
  modes}~\cite{zirn} fluctuating at scales of the order of the Fermi
wave length.  Choosing a representation in terms of eigenfunctions,
$\hat H |\psi_a\rangle = \epsilon_a |\psi_a\rangle$, wherein
$T=\{T^{\alpha\alpha'}_{a,a'}\}$ and $[\hat H,T] =
(\epsilon_a-\epsilon_a')\{T^{\alpha\alpha'}_{a,a'}\}$, we conclude
that there exist $N \sim \Eav /\Delta$ zero modes
$T_{\alpha\alpha'}^{aa'}$ whose action is controlled only by the
energy contribution, $S_{\rm qu}[T_{aa}]={i\beta\pi\omega\over 4\Eav}
\,{\rm tr}(\Lambda T_{aa}^{-1}\Lambda T_{aa})$. Upon Wigner
transformation, the zero mode operators turn into $N$ zero mode functions of
the action (\ref{eq:2}), rapidly fluctuating in classical phase
space. These modes are certainly not 'unphysical'. In the case of time
reversal non-invariant systems, one may indeed verify~\cite{nonnem} that
the integration over $T_{ab}$ leads to the formally exact eigenvalue
decomposition
\begin{equation}
  \label{eq:4}
R_2(\omega) = {\Delta^2 \over \Eav}
\sum_{|E_0-\epsilon_{a,b}|\le \Eav/2}
\delta(\omega-(\epsilon_a-\epsilon_b)).     
\end{equation}
What makes the rigorous derivation of Eq.~(\ref{eq:4}) possible is a
mathematical principle known as 'semiclassical exactness'.~\cite{fn5} (Incidentally, we
note that the possibility to obtain an exact representation of the
spectral correlation function from the quantum ballistic
$\sigma$-model shows that the approximations on which the derivation
of the latter is based -- saddle point approximation and first order
expansion in the quantum commutator -- become exact in the limit
$E_{\rm av}/\Delta\to \infty$.)

In spite of its formal correctness, Eq.~(\ref{eq:4}) is, of course,
useless in practice (much like the evaluation of the Feynman path
integral in an exact basis of eigenfunctions, formally an exact
alternative to the semiclassical stationary phase approximation, would be pointless.)

The conservative way to remove the zero modes is, again, by averaging
over weak randomness. Indeed, semiclassical exactness represents a
very delicate structure; even miniscule changes in the action will
spoil the exact cancellation of all non-Gaussian fluctuations on which
Eq.~\eqref{eq:4} is based. (Although we are lacking a rigorous
justification, we believe that averaging over a random potential whose
inverse scattering time, or level broadening, is as weak as $\hbar
\tau_q^{-1} \sim \Delta$ will suffice to effectively remove the zero modes.) It is, then, favorable to switch to a
Wigner phase space representation, and classify fluctuations along the
lines of the semiclassical analysis of section
\ref{sec:semiclassical-model}. Alternatively one may avoid averaging
over randomness and accept the presence of zero modes -- after all the
integration over these modes produces a meaningful, if useless
result. Mapping the integral onto a phase space representation
restricted to slowly fluctuating modes one may deliberately sacrifice the exact
information stored in the zero modes in return for a semiclassically
meaningful approximation scheme.

\bibliographystyle{apsrev}


\begin{thebibliography}{}


\bibitem{aleinerlarkin}
I. L.~Aleiner and A.~I.~Larkin, Phys. Rev. B {\bf 54}, 14423 (1996).

\bibitem{sieber_richter01}
M.~Sieber and K.~Richter, Physica Scripta {\bf T90}, 128 (2001); M.~Sieber, J. Phys. A {\bf 35}, L613 (2002).


\bibitem{smueller1}
S. M{\"{u}}ller, S. Heusler, P. Braun, F. Haake,  and A. Altland, Phys. Rev. Lett. {\bf 93}, 014103 (2004). 

\bibitem{smueller2}
S. M\"uller, S. Heusler, P. Braun, H. Haake, and A. Altland, Phys. Rev. E {\bf 72}, 046207 (2005).

\bibitem{smueller3}
S.~Heusler, S.~M\"uller, P.~Braun, and F.~Haake, J. Phys. A {\bf 37}, L31 (2004).


\bibitem{aa+haake}
S.~Heusler, S.~M\"uller, A.~Altland, P.~Braun, and F.~Haake, Phys. Rev. Lett. {\bf 98}, 044103 (2007).


\bibitem{efetov}
K.~B.~Efetov, Adv. Phys. {\bf 32}, 53 (1983).

\bibitem{bsm}
B.~A.~Muzykantskii and D.~E.~Khmel'nitskii, JETP Lett. {\bf 62}, 76 (1995).

\bibitem{ASAA1}
A.~V.~Andreev, B.~D.~Simons, O.~Agam, and B.~L.~Altshuler, Phys. Rev. Lett.  {\bf 76}, 3947 (1996).
\bibitem{ASAA2}
A.~V.~Andreev, B.~D.~Simons, O.~Agam, and B.~L.~Altshuler, Nucl. Phys. B {\bf 482}, 536 (1996).

\bibitem{BGS}
O.~Bohigas, M.~J.~Giannoni, and C.~Schmit, Phys. Rev. Lett. {\bf 52}, 1 (1984). 


\bibitem{fn1}
For definiteness, we
  assume that the quantum scale $\tE$ exceeds the time $t_{\rm mix}$
  after which the system dynamics becomes mixing. In the opposite
  case, the time range $t_E< t < t_{\rm mix}$ will be governed by a
  regime of irreversible, typically diffusive dynamics.



\bibitem{moyal}
J.~E.~Moyal, Proc. Cambridge Philos. Soc. Math. Phys. Sci {\bf 45}, 99 (1949). 

\bibitem{ruelle}
D.~Ruelle, Am. J. Math. {\bf 98}, 619 (1976).

\bibitem{kravtsovmirlin}
V.~E.~Kravtsov and A.~D.~Mirlin JETP-Lett. {\bf 60}, 645 (1994).

\bibitem{jmueller}
J.~M\"uller and A.~Altland, J. Phys. A: Math. Gen. {\bf 38}, 3097 (2005).


\bibitem{gaspard}
P.~Gaspard, {\it Chaos, Scattering and Statistical Mechanics} (Cambridge University Press, Cambridge, 1998). 


\bibitem{a1}
P.~W.~Brouwer, S.~Rahav, and C.~Tian, Phys. Rev. E {\bf 74}, 066208 (2006).

\bibitem{HOsum}
J.~H.~Hannay and A.~M.~Ozorio de Almeida, J. Phys. A {\bf 17}, 3429 (1984).


\bibitem{sc2}
D.~Spehner, J. Phys. A: Math Gen. {\bf 36}, 7269 (2003). 

\bibitem{fn2} 
For generic (time reversal non-invariant) systems, it
  is known~\cite{ruelle} that over a time range $\sim \lambda^{-1}
  (c^2/\epsilon)$ (the 'Ehrenfest' time corresponding to the initial
  distribution) $P_t$ propagates the initial distribution along the
  classical trajectory $\bx$, to then rapidly cross over to a uniform
  distribution $P_t\sim \Omega^{-1}$. We are not aware of equally
  rigorous results for time reversal invariant systems.

\bibitem{fn3}
Owing to their symmetry relations, the operators $T$
  are nonlinear objects which cannot be integrated, or Wigner
  transformed. What we mean by $T(\bx)$ is the function obtained by
  Wigner transformation of the (linear) generator $W$ in the
  exponential representation $T=\exp(W)$.


\bibitem{dis} J.~P.~M\"uller, {\it A Field Theory Approach to
    Universality in Quantum Chaos}, Ph.D. thesis, Universit\"at K\"oln
  (2007).



\bibitem{aleinerlarkin2}
I.~L.~Aleiner and A.~I.~Larkin, Phys. Rev. E {\bf 55}, R1243 (1997).


\bibitem{fn4}
The action-phase correction due to $V_q$ reads as
  $-{i\over \hbar}\int dt V_q({\bf q}(t))$. The (unperturbed classical) trajectories ${\bf q}$
  and ${\bf q'}$ effectively contributing to Gutzwillers double sum
  are classically close $|{\bf q}-{\bf q'}|\sim \hbar$, implying a
  vanishing of the phase differences $-{i\over \hbar}\int dt (V_q({\bf
    q}(t))-V_q({\bf q'}(t)))$ in the semiclassical limit.


\bibitem{zirn}
M.~R.~Zirnbauer, in {\it Supersymmetrie and Trace Formulae: Chaos and Disorder,} ed. by I.~V.~Lerner, J.~P.~Keating, and D.~E.~Khmel'nitskii (Plenum Press, New York, 1999), p.~17. 

\bibitem{nonnem}
S.~Nonnemacher and M.~Zirnbauer,  J. Math. Phys. {\bf 43}, 2214 (2002).

\bibitem{fn5}
It is a well known mathematical fact that integrals over certain
  manifolds (manifolds possessing a symplectic structure) may be
  exactly evaluated by stationary phase methods. Finite dimensional
  applications of this principle include the Itzykson-Zuber
  formulae~\cite{s1}, and the mean field approach to spectral
  correlations for certain symmetry classes~\cite{andreevaltshuler}.
  The ballistic $\sigma$-model projected onto an exact
  eigenbasis can be treated by the same methods.


\bibitem{s1}
C. Itzykson and J. B. Zuber, J. Math. Phys. {\bf 21}, 411 (1980).

\bibitem{andreevaltshuler}
A. V. Andreev and B. L. Altshuler, Phys. Rev. Lett. {\bf 75}, 902 (1995).

\end{thebibliography}

\end{document}